# The QIC-Index: A Novel, Data-Centric Metric for Quantifying the Impact of Research Data Sharing

**Martin G. Frasch**[1]
University of Washington, Seattle, WA, USA[2]

## Abstract

Modern science faces an "incentive gap" where traditional, publication-centric metrics like the h-index fail to value the critical contribution of research data sharing. This discourages open science practices and hinders collaborative progress. To address this, we propose the QIC-Index, a novel metric designed to quantify and reward the sharing of high-quality research data. The QIC-Index moves beyond publications to assess the value of individual data objects. The score for each shared data object ($s_j$) is calculated as a product of its **Quality** ($Q_j$), **Impact** ($I_j$), and **Collaboration** ($C_j$) scores. An author's total QIC-Index ($S_i$) is the sum of these individual scores across all their shared data contributions ($S_i = \Sigma\ s_j$). This framework incentivizes the sharing of data that is not only high-quality and impactful but also the product of meaningful teamwork. By aligning individual rewards with the collective goals of open science, the QIC-Index offers a robust tool to foster a more transparent, efficient, and collaborative research culture.

---

[1] Contact: mfrasch@uw.edu or martin@healthstreamanalytics.com
[2] Originally submitted to the NIH S-Index challenge 06/01/2025

# 1. Introduction: The Incentive Gap in Research Evaluation

The current academic reward system, dominated by metrics like the h-index [1], primarily values peer-reviewed publications. This creates a structural inability to recognize or reward non-publication outputs, such as the creation and sharing of high-quality, reusable datasets [2]. This "incentive gap" discourages the foundational work of data sharing, which is essential for reproducibility, innovation, and large-scale "team science" [3]. Major policy initiatives, including the NIH's 2023 Data Management and Sharing Policy, have highlighted the urgent need for new metrics that can assess the quality and impact of data sharing to drive a necessary cultural transformation [4].

Several approaches have been proposed to address this gap. One notable effort is the S-Index developed by Olfson, Wall, and Blanco (2017), which applies an h-index-like calculation to the set of publications that *use* an investigator's shared data [5]. While this was a significant step toward valuing data, the Olfson S-Index remains an indirect, publication-centric measure; a high-quality dataset is only valued if it is reused in a highly-cited paper.

This paper introduces the **QIC-Index**, a metric that shifts the unit of analysis from the publication to the data object itself. It provides a more direct, holistic, and data-centric measure by evaluating the intrinsic quality, actual reuse, and collaborative context of the shared data. The following table compares these three metrics:

Table 1: Comparative Analysis of Data-Centric Impact Metrics

| Attribute | H-Index | Olfson S-Index (2017) | QIC-Index (Proposed) |
|---|---|---|---|
| Primary Unit of Analysis | Author's Publications | Publications *Citing* Shared Data | Shared Data Object (Dataset/Software) |
| Core Principle | Balances publication quantity & citation impact. | Measures citation impact of research that *reuses* shared data. | Measures intrinsic quality (FAIR), reuse impact, and collaborative context of the data object. |
| Key Data Inputs | Publication citation network. | Data citation links; publication citation network. | Repository metadata (APIs); data reuse network; author affiliations. |

| Primary Limitation | Blind to non-publication outputs. | Indirect measure; value is tied to downstream publication success. | High dependency on immature data infrastructure; complex calculations. |

## 2. The QIC-Index Framework

The QIC-Index for an individual researcher ($S_i$) is the sum of the scores of all data objects ($s_j$) they have contributed to:

$$S_i = \Sigma\, s_j$$

The score for each data object is a product of three components:

$$s_j = Q_j \times I_j \times C_j$$

This multiplicative design ensures a balanced contribution. A dataset of exceptional quality ($Q_j$ is high) that is never used ($I_j$ is zero) contributes nothing to the index.

## 3. Methodology: Quantifying Quality, Impact, and Collaboration

### 3.1 The Quality (Q) Score

The Quality score operationalizes the FAIR Guiding Principles (Findable, Accessible, Interoperable, Reusable) into a quantitative measure [6]. It is a weighted average of four sub-scores:

$$Q_j = w_f\, q(F,j) + w_a\, q(A,j) + w_i\, q(I,j) + w_r\, q(R,j)$$

We acknowledge the technical challenge in fully automating this calculation due to heterogeneous repository metadata. Therefore, an initial, practical deployment of the QIC-Index would likely rely on a semi-automated approach, where algorithms provide a preliminary score that is verified by human curators.

### 3.2 The Impact (I) Score

The Impact score measures the actual scientific utilization of a shared data object. The formula captures meaningful reuse and tempers the effect of single outliers through a logarithmic scale:

$$I_j = 1 + \ln(1 + \Sigma\, w(reuse,k))\ ^3$$

This calculation is dependent on a robust data citation and tracking infrastructure, which is still nascent. Initiatives like **Make Data Count** are developing the community standards and open infrastructure needed to track data usage and citation more systematically [7]. The QIC-Index is designed not merely to consume this data, but to act as a catalyst to drive the adoption of these better data citation practices by creating a tangible reward.

---

[3] The summation is performed over all M reuse events, from k=1 to M.

### 3.3 The Collaboration (C) Score

The Collaboration score directly rewards collaborative science by quantifying the breadth of the network involved in creating a data object [8]:

$$C_j = (1 + \ln(N_{authors})) \times (1 + 0.5 \times \ln(N_{institutions}))$$

The formula incorporates both the number of authors and institutions, with the 0.5 coefficient providing a significant boost for multi-institution efforts.

## 4. System Architecture and Implementation

The calculation of the QIC-Index relies on a robust data pipeline and knowledge graph. The workflow ingests new data contributions, analyzes them, and continuously monitors their impact over time.

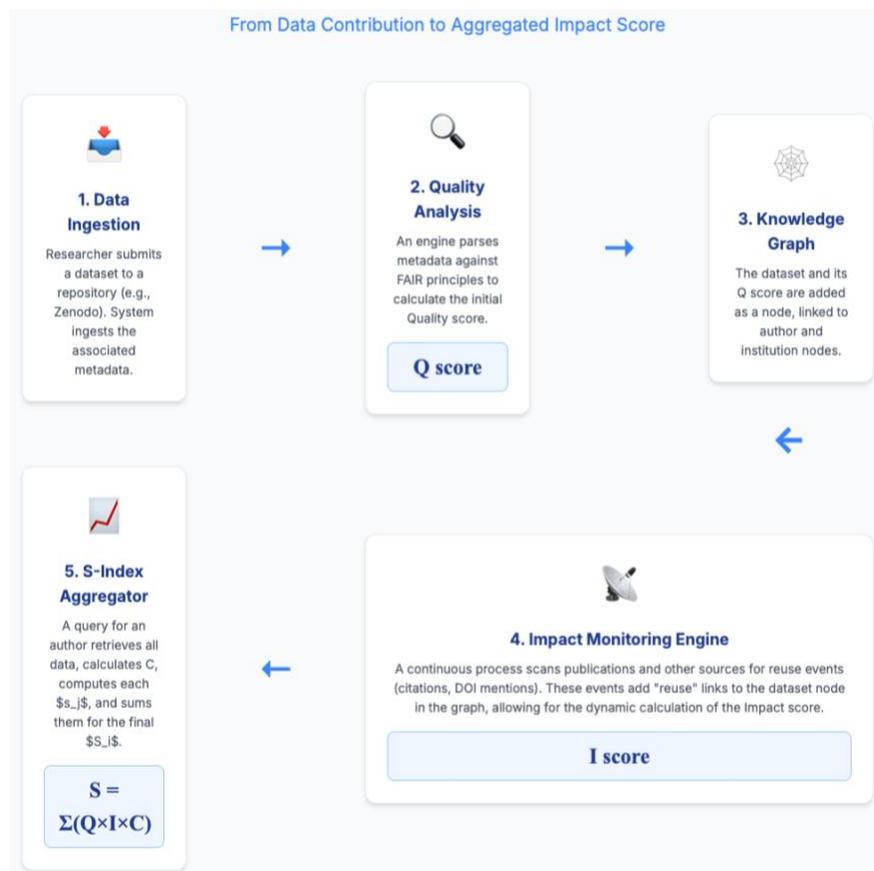

Figure 1. The S-Index System Architecture (Q*I*C Framework).

The process flows as follows: Data Ingestion → Quality Analysis Engine (Q Score) → Knowledge Graph Integration → Impact Monitoring Engine (I Score) → QIC-Index Aggregator (C Score and Final Calculation).

## 5. Comparative Analysis: QIC-Index in Practice

To illustrate the practical implications of the QIC-Index, we present two personas:

- **Dr. Singh (The Traditional PI):** A highly-cited researcher with an h-index of 60. She has published extensively but has shared few datasets. Her **QIC-Index is 45**. Her score is derived primarily from one major, meticulously curated longitudinal dataset that, while only supporting a moderate number of her own publications, has been reused extensively by the broader community.
- **Dr. Al-Jamil (The Early-Career Collaborator):** An early-career researcher with a modest h-index of 8. She is a key contributor to a large, multi-institution consortium. Her **QIC-Index is 12**. Her score reflects her significant contributions to two large consortium datasets that have high Quality and Collaboration scores, rewarding her teamwork even before the datasets have accrued significant reuse.

While the h-index suggests Dr. Singh is vastly more impactful, the QIC-Index provides a more nuanced view, recognizing the significant, yet traditionally invisible, contributions of Dr. Al-Jamil.

## 6. Conclusion: A Call to Action

The QIC-Index is a purpose-built tool designed to make the vital contributions of data sharing both visible and valuable.[4] It is not intended to replace existing metrics, but rather to complement them and address known gaps in research evaluation [9]. By adopting tools like the QIC-Index, we can begin to build a research culture that is more collaborative, transparent, and equipped to solve the scientific challenges of the 21st century.

---

[4] GitHub repository: https://github.com/martinfrasch/s-index